\documentclass{WileyMSP-template}
\usepackage{ragged2e}
\usepackage[colorlinks=true,linkcolor=blue,citecolor=blue,urlcolor=black]{hyperref}

\begin{document}

\pagestyle{fancy}
\rhead{\includegraphics[width=2.5cm]{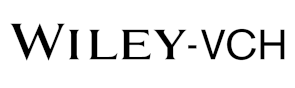}}

\title{Colloidal Nanocrystals Regrowth-Assisted Synthesis of Perovskite Microwire Lasers for Integrated Optoelectronics}

\maketitle

\author{Elizaveta V. Sapozhnikova\textsuperscript{1,2$\dagger$},}
\author{Ivan A. Matchenya\textsuperscript{1$\dagger$},}
\author{Dmitry A. Tatarinov\textsuperscript{1,2},}
\author{Grigorii A. Verkhogliadov\textsuperscript{1},}
\author{Dmitry A. Semyonov\textsuperscript{2},}
\author{Maria A. Kirsanova\textsuperscript{1},}
\author{Natalia K. Kuzmenko\textsuperscript{2},}
\author{Julia S. Mironova\textsuperscript{1},}
\author{Arina O. Kalganova\textsuperscript{1},}
\author{Valeriya M. Levkovskaya\textsuperscript{1},}
\author{Stepan A. Baryshev\textsuperscript{1},}
\author{Yuxi Tian\textsuperscript{3},}
\author{Anatoly P. Pushkarev\textsuperscript{1}*}

\begin{affiliations}

$^1$ Skolkovo Institute of Science and Technology, Bolshoy Boulevard 30, bldg. 1, Moscow, 121205, Russia.

$^2$ ITMO University, Kronverksky Pr. 49, bldg. A, St. Petersburg, 197101, Russia.

$^3$ Key Laboratory of Mesoscopic Chemistry of MOE, State Key Laboratory of Analytical Chemistry for Life Science, School of Chemistry and Chemical Engineering, Nanjing University, 210023 Nanjing, China.

$^{\dagger}$These authors contributed equally to this work.

Anatoly P. Pushkarev

Email Address: an.pushkarev@skoltech.ru

\end{affiliations}


\begin{abstract}
\justifying 
Colloidal perovskite nanocrystals (NCs) are a well-proven platform for growing anisotropic structures. Nanowires (NWs) exhibiting a quantum confinement phenomenon and microwires (MWs), which enable lasing, are of particular interest for optoelectronic devices. Synthesis of the latter is challenging. Herein, we report a straightforward access to high-quality CsPbBr$_3$ MW lasers. We utilize a diphenyl ether (DPE) solvent for the hot-injection synthesis. DPE coordinates strongly to Pb$^{2+}$ and allows to reduce an excess of oleic acid/oleylamine ligand pair well established for PbBr$_2$ dissolution and inhibition of as-formed NCs regrowth. Therefore, a rapid injection of Cs-oleate into the PbBr$_2$-containing solution yields lead-depleted Cs$_4$PbBr$_6$ NCs which slowly release perovskite precursors and produce CsPbBr$_3$ counterparts. The latter transform into NWs through an oriented-attachment mechanism, which in turn evolve into laser MWs. To demonstrate spectrally tunable lasing in MWs we employ YCl$_3$ for ion exchange in perovskite lattice. Resultant CsPb(Cl,Br)$_3$ MWs show high-Q coherent emission in the 485--540 nm range. To highlight the potential of synthesized MWs for integrated optoelectronics, we assemble a device comprising a CsPb(Cl,Br)$_3$ MW laser coupled to MoO$_3$ lossless nanowaveguide, which delivers coherent light to a CsPbBr$_3$ MW photodetector. The device exhibits a nonlinear optoelectronic response applicable for on-chip neuromorphic computing.

\end{abstract}


\section{Introduction}
\justifying 
Over the last decade, halide perovskite nanocrystals (NCs) have emerged as brightly luminescent materials with tunable bandgap for advanced photonics and optoelectronics~\cite{dey2021state}. Owing to electronic structure of halide perovskites which defines their ``defect tolerance"~\cite{brandt2015identifying} (i.e. formation of shallow defect states only near the valence band maximum and conduction band minimum), passivation of surface traps by using organic ligands~\cite{pan2017bidentate,zhang2019alkyl,imran2019simultaneous,grisorio2022situ}, and relatively large exciton binding energies~\cite{protesescu2015nanocrystals} colloidal solutions of NCs exhibit narrow-band emission with near-unity photoluminescence quantum yield (PLQY) at room temperature. These properties along with large light absorption coefficient~\cite{maes2018light} and high charge carrier mobility in perovskites~\cite{herz2017charge} promote NCs for the fabrication of high-performance light-emitting diodes~\cite{wang2020dimension,zheng2021cspbbr3,guo2023highly}, solar cells~\cite{zhang2020alpha,hu2021flexible}, and photodetectors~\cite{kang2019high,liu2023high}. In the meanwhile, all-inorganic perovskite NCs represent a promising platform for the synthesis of one-dimensional structures and their self-assembled clusters~\cite{imran2016colloidal,tong2017precursor,liu2019light,hudait2020facets,pradhan2021halide,behera2023hexahedron,bi2020self}.

Various colloidal synthetic routes have been reported for thin CsPbX$_3$ (X = Cl, Br, I) nanowires (NWs). A standard air-free technique which utilizes alkyl amine ligands (octylamine$-$OctAm and oleylamine$-$OlAm) only was shown to give CsPbBr$_3$ NWs with 10$-$20 nm width, whereas introducing carboxylic acids with short aliphatic chains (octanoic acid$-$OctAc or hexanoic one$-$HexAc) into the reaction mixture invoked the formation of thinner NWs (3.4$-$5.1 nm width) demonstrating tunable absorption and photoluminescence (PL) owing to quantum confinement regime~\cite{imran2016colloidal}. Liu et al.~\cite{liu2019light} revealed that an excess of OlAm in the standard reaction~\cite{protesescu2015nanocrystals} facilitates oleylammonium$-$oleate (OlAm$-$OA) pairs release from CsPbBr$_3$ NCs surface under visible light irradiation. Hence, non-passivated highly reactive CsBr- and PbBr$_2$-terminated (110) facets of orthorhombic NCs trigger the anisotropic assembly~\cite{liu2019light}. Similar oriented-attachment mechanism was established for the ligand-assisted direct conversion of Cs$_2$CO$_3$ and PbX$_2$ precursor powders into colloidal NWs (ca. 12 nm in width) by ultrasonication~\cite{tong2017precursor}. Along with small cubic-shaped NCs regrowth that occurs in [001], [1-10]~\cite{imran2016colloidal}, and [110]~\cite{liu2019light} directions, there is also possibility to obtain structures elongated in [010] or [100] ones~\cite{hudait2020facets}. Producing such zigzag-shaped 1D structures is attributed to Pb$^{2+}$-rich active (100) and (112) facets of large polyhedral CsPbBr$_3$ NCs (size of 35$-$50 nm) that afford corner-wise connections~\cite{hudait2020facets}. Importantly, quantum confined NWs capable of self-assembly into robust superstructures with improved photo- and environmental stability. Thus, CsPbI$_3$ NWs clusters emitting at 600 nm were employed for high-performance perovskite LEDs~\cite{bi2020self}.  

A transition from CsPbX$_3$ nanowires to bulk microwires possessing the subwavelength cross-section dramatically reduces Auger recombination in semiconductor medium~\cite{garcia2009suppressed}. This, along with high optical gain (above 10 000 cm$^{-1}$ for CsPbBr$_3$~\cite{tatarinov2023high}), moderate refractive index (n~$\approx$~2.5), and well reflective end facets promotes perovskite MWs for low-threshold high-Q Fabry-P\'{e}rot (FP) lasers~\cite{zhu2015lead,zhou2017vapor,markina2023perovskite}. Colloidal synthesis of MW lasers is challenging since the approach being developed has to address the following issues: polycrystalline structure, irregular shape of laser cavities, and their coalescence in solution. Combining colloidal MW lasers and state-of-the-art fabrication techniques, e.g., micro pick-and-place, optical tweezers, it would be much easier to assemble various simple designs~\cite{zhao2018switchable,marunchenko2023mixed,matchenya2025short} as well as entire circuits~\cite{han2024inorganic}. Therefore, resolving the aforementioned issues is a prerequisite for large-scale inexpensive production of photonic and optoelectronic integrated devices.

In this work, we propose a novel approach to the colloidal synthesis of MW lasers. In the standard reaction for CsPbBr$_3$ NCs~\cite{protesescu2015nanocrystals}, we change a commonly used 1-octadecene (ODE) solvent to diphenyl ether (DPE) which coordinates to Pb$^{2+}$ and allows us to obtain a clear solution for PbBr$_2$, OA, and OlAm mixture without a large excess of the ligands at 140 $^{\circ}$C. Injection of Cs-oleate (CsOA) into the solution yields orthorhombic CsPbBr$_3$ small NCs along with sacrificial rhombohedral Cs$_4$PbBr$_6$ large NCs. Prolonged incubation at 140 $^{\circ}$C slowly dissociates the sacrificial phase and gives origin to the middle-sized CsPbBr$_3$ NCs~$-$~building blocks of thin NWs. Further heating of the solution at 180 $^{\circ}$C for 1 h transforms NCs into MWs. By taking aliquots at the incubation stages and examining their content, in particular size, shape, and crystal phase of particles, using transmission electron microscopy, fluorescence microscopy, and X-ray diffraction, respectively, we straightforwardly reconstruct the entire colloidal evolution that has been elusive for numerous studies~\cite{pradhan2022growth}. As-grown MWs demonstrate high crystallinity and properly chipped end facet,s which enable optical eigenmodes in a FP cavity. It is established that MWs transferred to DPE:hexane solution do not coalesce. Furthermore, halide exchange-induced tuning of a spontaneous emission peak from 525 nm to 485 nm can be addressed by introducing an YCl$_3$ additive into the DPE:hexane solution. Pristine CsPbBr$_3$ and mixed-halide CsPb(Cl,Br)$_3$ MWs exhibit low-threshold high-Q lasing under femtosecond laser pulses. Finally, we employ MWs of different compositions for assembling a compact integrated device. In this device, coherent light leaks from CsPb(Cl,Br)$_3$ MW laser into optically transparent MoO$_3$ nanowaveguide, propagates along it at a distance of ca. 84 $\upmu$m, decouples from the opposite end facet, and excites CsPbBr$_3$ MW photodetector. Nonlinear optoelectronic response of the device is discussed.

\section{Results and Discussion}
\label{results_discussion}

\subsection{Synthesis of CsPbBr$_3$ microwires}
Our synthetic approach includes two major stages. At the first stage, we modify a protocol by Protesescu et al.~\cite{protesescu2015nanocrystals}. In particular, we use tow times smaller amount of PbBr$_2$ (0.09 mmol), three times smaller amount of OA and OlAm (0.46 mmol) ligands, and diphenyl ether (DPE, 7 mL) solvent in lieu of 1-octadecene (ODE, 5 mL). PbBr$_2$ powder is vacuumized in DPE under stirring (400 rpm) before the injection of ligands at 120 $^\circ$C for 1 h. Then, the flask is filled with N$_2$ gas and ligands are introduced into the solution. Thereafter, the temperature of the mixture increases up to 140 $^{\circ}$C and the PbBr$_2$ powder dissolves completely to give a clear colorless solution that is quite surprising because of the reduced excess of ligands taken for the reaction. We assume this happens owing to DPE solvent coordinates strongly to Pb$^{2+}$~\cite{chiang2025investigating} and hence allows us to utilize fewer ligands that are recognized to facilitate the dissolution of the PbBr$_2$ powder and, to some extent, prevent the coalescence and growth of as-formed NCs. Freshly prepared CsOA (0.05 mmol) in ODE (0.125 M, for synthesis see Experimental Section) is swiftly injected into the solution at 140 $^{\circ}$C that results in it turning a pale yellow hue. We incubate this solution at 140 $^{\circ}$C for 2 h and observe the gradual change in its color to a bright yellow one. At the second stage, the solution is heated up to 180 $^{\circ}$C and stirred for 1 h more to obtain a desired product which has a dark orange color. To investigate the entire evolution of colloidal particles, aliquots (0.2 mL) are taken at both stages. The content of these aliquots is studied using X-ray diffraction, fluorescence microscopy, and transmission electron microscopy for exploring the evolution on the macro-, micro- and nanoscale. Finally, the solution with the target product is quenched by using an ice bath. The described synthesis is schematically illustrated in Figure 1.    

\begin{figure*}[t!]
\centering
\includegraphics[width=0.66 \columnwidth]{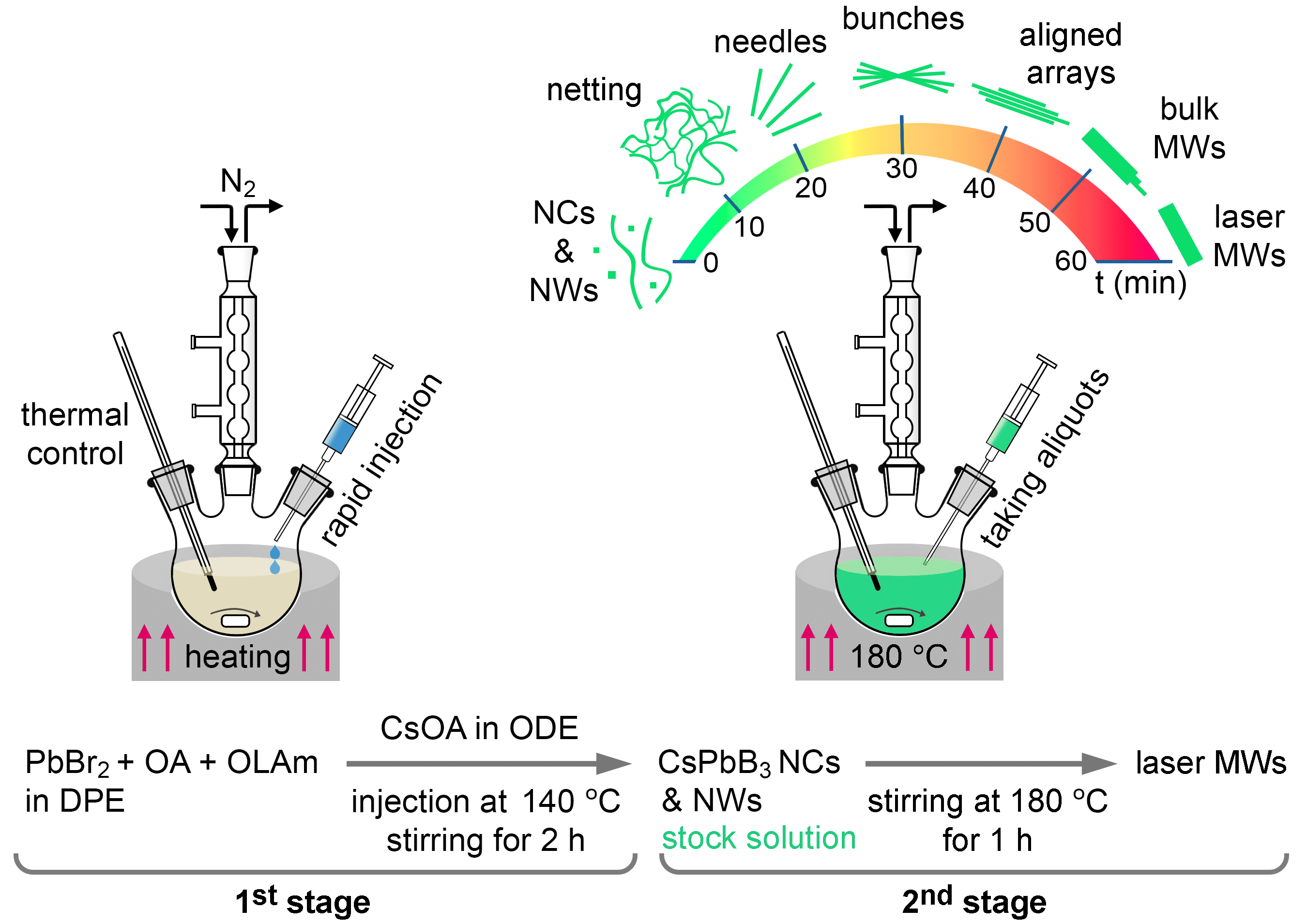}
\caption{Schematic illustration for the synthesis of laser MWs. Injection of CsOA into PbBr$_2$/ligands solution at 140 $^o$C and stirring the formed colloid for 2 h gives a stock solution with NCs and NWs. Heating the stock solution up to 180 $^o$C at constant stirring and taking aliquots every 10 min makes it possible to track the evolution of colloidal particles.}
\label{fig1}
\end{figure*}

\subsection{Structural characterization of colloidal particles}

\begin{figure*}[h!]
\centering
\includegraphics[width=1\columnwidth]{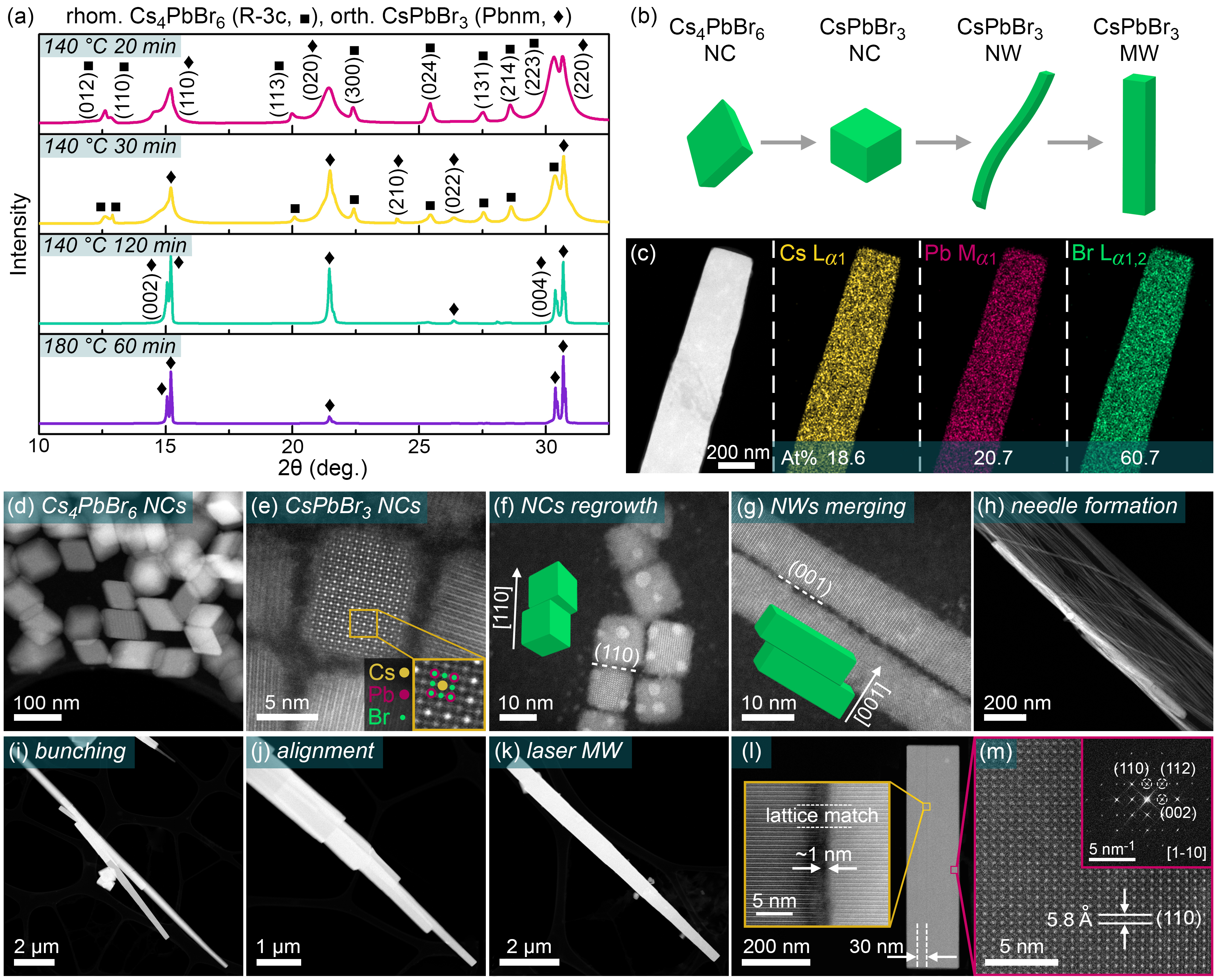}
\caption{(a) XRD patterns of colloidal particles from aliquots collected at the first (red, yellow, and green lines) and second (purple line) stages of the synthesis. Square and rhombus symbols identify diffraction peaks belonging to rhombohedral Cs$_4$PbBr$_6$ (space group R-3c) NCs and orthorhombic CsPbBr$_3$ (space group Pbnm) particles, respectively. Three panels (from top to bottom) show the transformation of the non-perovskite phase to the perovskite one during the incubation at the first stage. The bottom panel shows an increase in crystallinity of 1D anisotropic CsPbBr$_3$ particles presented by doublets with major peaks (110) and (220) in the diffraction pattern of the product obtained at the end of the second stage. (b) Schematic illustration of colloidal particles evolution. (c) EDX mapping of a single MW showing an even spatial distribution of Cs, Pb, and Br atoms in the crystal lattice and nearly stoichiometric chemical content. (d-g) HAADF-STEM images of the particles taken at the first stage. Large-sized Cs$_4$PbBr$_6$ NCs (d) serve as a source of precursor species for middle- [7 nm, (e)] and standard-sized [11-12 nm, (f)] CsPbBr$_3$ NCs. Atomically resolved image of a single middle-sized NC (e) allows for determining positions of Cs, Pb, and Br atoms in the crystal lattice [inset image in (e)]. Standard-sized NCs regrow along [110] direction (f) and yield NWs which attach each other through (001) faces (g). (h-k) Images of the particles taken at the second stage. Arrays of NWs form needle-like structures (h), which assemble bunches (i). Bunches undergo internal alignment (j) and chipping of the uneven ends to give Fabry-P\'{e}rot laser microcavities (microwires) with reflective end facets (k). (l,m) High-resolution images visualizing a small gap between lattice-matched 30 nm thick NWs (boundaries are highlighted with dashed lines) within a single microcavity [inset image in (l)]. Selected area image and its FFT confirm high crystallinity of the microcavity.}
\label{fig2}
\end{figure*}

To characterize the crystal structure of the product that forms right after the injection of CsOA and its possible transformation, which is, apparently, reflected in the change of the solution color, sediments from the aliquots taken at the first stage of the synthesis are examined by X-ray diffraction (XRD). We notice the color of the initially pale yellow solution is getting brighter in 30 min after the injection. In accordance with these observations, the diffraction patterns of aliquots taken within 20 min (top panel in Figure 2a) look very similar and consist of peaks (012), (110), (113), (300), (024), (131), (214), and (223) belonging to rhombohedral Cs$_4$PbBr$_6$ (space group R-3c~\cite{velazquez2008growth}) along with broad signals (110), (020), and (220) of orthorhombic CsPbBr$_3$ (space group Pbnm~\cite{rodova2003phase}). The reason for broad peaks could be small-sized CsPbBr$_3$ particles. When the solution is incubated for 30 min, peaks assigned to the rhombohedral phase go down and completely vanish as the incubation time reaches 120 min (middle panels in Figure 2a). On the contrary, peaks corresponding to the orthorhombic phase are getting narrower and more intensive that implies improvement of crystallinity of colloidal particles, i.e. increase in their dimensions. We then compare the sediment from the solution kept at 140 $^{\circ}$C for 2 h (the end of the first stage) and final product obtained by heating at 180 $^{\circ}$C for 1 h (bottom panel in Figure 2a). The former exhibits a pronounced diffraction peak (020) at ca. 21.4$^{\circ}$ 2$\theta$, whilst the latter demonstrates doublets at ca. 15.1$^{\circ}$ and 30.5$^{\circ}$ 2$\theta$ mostly. The dominant peaks (110) and (220) in the bottom-panel pattern in Figure 2a confirm the growth of highly crystalline CsPbBr$_3$ microwires along [110] direction at the second stage of the synthesis. The transformation of the colloidal particles is illustrated in Figure 2b. The resultant product is measured by energy dispersive X-ray spectroscopy (EDX) that confirms a uniform distribution of all the elements in the volume of a single microwire and its chemical content Cs:Pb:Br 18.6:20.7:60.7 At$\%$ close to that of stoichiometric perovskite (Figure 2c).

Complementary to XRD, colloidal particles extracted from the aliquots taken at both stages are measured by high-angle annular dark-field scanning transmission electron microscopy (HAADF-STEM). The images show large-sized Cs$_4$PbBr$_6$ NCs (Figure 2d) along with small CsPbBr$_3$ NCs (Figure S1) that form right after the CsOA injection. Middle-sized perovskite NCs of ca. 7 nm can be found in the 30 min aliquot (Figure 2e). High-resolution image of a single NC allows to identify atomic positions for perovskite lattice (inset in Figure 2e). The incubation of the solution at 140 $^{\circ}$C for 2 h gives NCs of 11-12 nm size (Figure 2f) which undergo
anisotropic linear growth along [110] direction via oriented attachment and yield NWs (Figure 2g). In turn, NWs attach to each other through (001) surfaces. Importantly, no Cs$_4$PbBr$_6$ NCs are found in 120 min aliquot. Therefore, the rhombohedral phase appears to be sacrificial and most likely dissociates into perovskite precursor species, contributing to the growth of CsPbBr$_3$ NCs.

The second stage aliquots taken subsequently over 60 min of heating the solution at 180 $^{\circ}$C reveal the evolution of anisotropic particles mainly. It is established that merging of NWs occurs and needle-like structures form within arrays of long uniform NWs (Figure 2h). Then, needles tend to combine into non-aligned bunches (Figure 2i). Crystallization and alignment of the needles within bunches produce microwires with uneven ends (Figure 2j). Such ends are not able to reflect internal light traveling along the microwire, therefore, a standing wave (optical eigenmode) cannot exist in these conditions. However, chipping the uneven ends naturally provides a Fabry-P\'{e}rot cavity with good reflective facets (Figure 2k). A high-resolution HAADF-STEM image of an isolated regularly-shaped microcavity visualizes some of its constituents $-$ NWs of 30 nm width (dashed lines in Figure 2l are guides to the eye). Note there is good matching between two crystal lattices of the constituents, although they are separated by a 1 nm gap (inset image in Figure 2l). Except these rare intracavity boundaries, the microcrystal itself demonstrates excellent crystallinity and the distance of 0.58 nm between the (110) crystallographic planes proven by atomically resolved image (Figure 2m) and its fast Fourier transform (FFT, inset image in Figure 2m).

To gain deeper insights into the synthesis of laser MWs, we compare our approach with light-induced self-assembly of cubic CsPbBr$_3$ perovskite NCs into NWs and assembly of thin NWs into thick NWs that happens via oriented attachment through (110) and (001) surfaces, respectively~\cite{liu2019light}. To exclude the possible impact of light-induced assembly on the evolution of colloidal particles, we conduct the synthesis in dark conditions and obtain the same resultant product. A major advantage of our approach over the reported ones elsewhere is the formation of separate MWs in the solution instead of agglomerated NWs ~\cite{liu2019light, imran2016colloidal, tong2017precursor}. A slow release of perovskite precursor species by sacrificial lead-depleted Cs$_4$PbBr$_6$ NCs enables the formation of CsPbBr$_3$ of the required size gradually over an extended period. That is why these NCs have a chance to develop into very long NWs. After arrays of NWs form needles, the latter experience adhesive van der Waals forces and undergo bunching. The alignment of the needles within bunches occurs owing to van der Waals-London dispersion torque~\cite{rajter2007van}. Finally, we emphasize the role of stirring for obtaining the target product. Stirring invokes high shear forces that break long arrays of NWs into short ones and assist the formation of separate needles. Furthermore, stirring causes collisions of microscopic particles that is supposed to be the main reason for chipping MWs' uneven ends. 

Remarkably, a parallel synthetic route also takes place, leading to the formation of MWs with poor crystal quality. This route slightly impedes the synthesis of laser MWs. It involves the attachment of cubic NCs or short NWs to long NWs, together with the coalescence of long NWs, resulting in 1D structures with rough morphology and voids. (Figure S2). 

\begin{figure*}[t!]
\centering
\includegraphics[width=1\columnwidth]{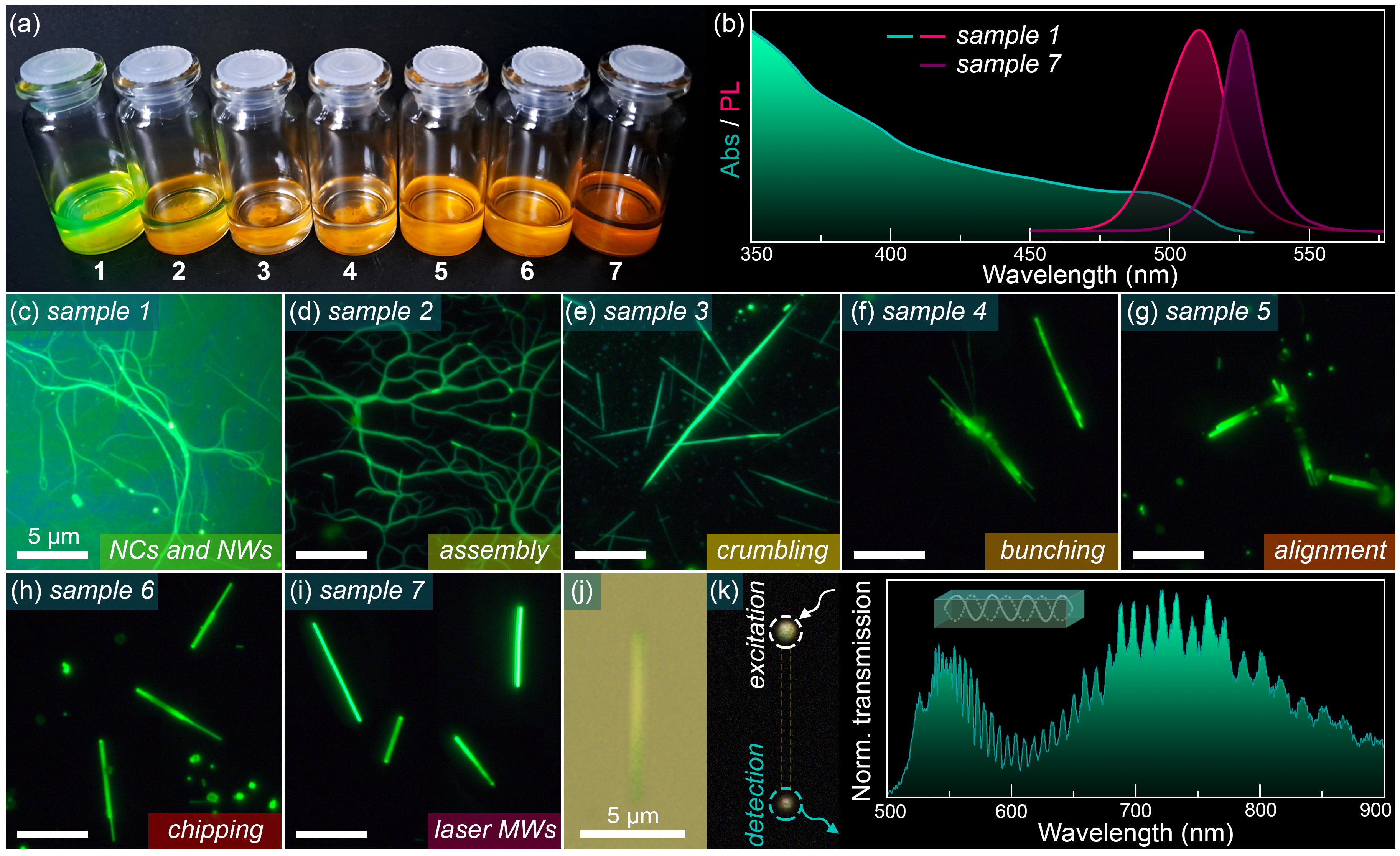}
\caption{(a) A picture of the aliquots (samples) \textbf{1}-\textbf{7} diluted in hexane, where \textbf{1} is collected at the beginning of the second stage, and \textbf{7} is the target product. Greenish fluorescence in sample 1 stems from colloidal NCs, whereas bright yellow sediment is NWs arrays. The size of particles in the aliquots increases from left to right that manifests in the gradual change in the sediment color. The dark orange color of the sample 7 is caused by substantial light absorption in bulky CsPbBr$_3$ MWs. (b) Optical absorption and PL spectra for \textbf{1} and \textbf{7}. (c-i) Fluorescence microimages illustrating the stepwise synthetic formation of laser MWs, including the key steps: assembly of NWs arrays into netting, its crumbling into needles, bunching of the needles, their alignment within the bunches, and chipping of MWs with uneven ends. (j) Bright-field image of an isolated laser MW. (k) Dark-field image of the same MW excited by focused white light at the end facet and spectrum of transmitted light collected from the opposite end. Modulation in the spectrum is caused by Fabry-P\'{e}rot optical modes, indicating that reflective end facets of the MW enable standing waves [inset picture in (k)].}
\label{fig3}
\end{figure*}

\subsection{Optical imaging of the particles}

With the inspiring result at hand, we carry out a basic optical characterization of the colloidal particles and track their evolution on the microscale using fluorescence microscopy. The second stage aliquots (samples) under investigation are presented in Figure 3a:  1$^{st}$ is collected right after the heating up to 180 $^o$C, and the rest of aliquots are taken every 10 min over the incubation procedure (Figure 1). One can see how the color of the sediment changes from bright yellow to dark orange that is related to enhanced light absorption by microcrystals as compared to that of nanocrystals. The transition from NCs, exhibiting quantum confinement phenomenon, to bulk perovskite is clearly observed in absorption and photoluminescence (PL) spectra (Figure 3b). Sample \textbf{1} demonstrates broad PL spectrum (FWHM = 26 nm, full width at half maximum) peaked at 510 nm and a Stokes shift of 19 nm due to wide particle size distribution. Sample \textbf{7}, by contrast, contains large particles only (mainly MWs). Therefore, its absorption spectrum cannot be measured, whilst the PL signal has a maximum at 525 nm and FWHM = 16 nm.

Fluorescence microimages of the samples \textbf{1}-\textbf{7} deposited on glass slides (Figure 3c-i) are consistent with HAADF-STEM ones and prove the above proposed scenario: i) CsPbBr$_3$ NCs and arrays of long NWs coexist at the beginning of the second stage; ii) then, all the NCs contribute to the extensive growth of NWs arrays which, in turn, connect each other and form a netting; iii) the netting crumbles into numerous needles; iv) the needles combine into bunches; v) thereafter, the needles align within the bunches; vi) this alignment yields MWs with uneven ends; vii) finally, chipping of the uneven ends results in laser MWs of 5--20 $\upmu$m length and possessing the subwavelength cross-section.

To visualize the absence of submicron crystalline inclusions in laser microwires, we demonstrate bright-field (BF, Figure 3j) and dark-field (DF, Figure 3k) optical images of a single MW. For the DF image, focused white light excites the MW at one of the end facets. One can see the light confined inside the cavity outcouples exclusively from the end facets and does not scatter while traveling along the microcavity~\cite{pushkarev2019few}. A spectrum of the transmitted light collected from the opposite end reveals FP modulation with uneven mode spacing that becomes smaller as FP modes approach the exciton resonance (inset plot in Figure 3k). Such a trend is a signature of exciton-polaritons (EPs) occupying discrete energy states in the low polariton branch (LPB) of CsPbBr$_3$ microcavity~\cite{shang2020role}.  

\subsection{Composition tunable lasing in microwires}

Optical properties of halide perovskites can be precisely regulated by postsynthesis composition engineering of ABX$_3$ lattice that occurs through vacancy-mediated ion exchange. The rate of the ion exchange is mostly affected by the activation energy of the A--, B--, and X--sites, where the latter one exhibits the lowest energy value~\cite{eames2015ionic}. Therefore, the rapid and complete halide exchange taking place in reactions between certain halide salts~\cite{akkerman2015tuning,ramasamy2016all,zhang2017full,yan2018tuning,tong2023situ,livakas2023cspbcl3,huang2025surface} dissolved in solvents and perovskite-dispersed solutions is recognized to be the most efficient strategy towards tuning the optical bandgap. However, there are two drawbacks of such liquid-state reactions: i) the entire halide exchange in NCs can be completed within a short interval (minutes or seconds) that complicates fine control over the reaction; ii) the exchange reduces PLQY of the alloyed NCs~\cite{akkerman2015tuning,zhang2017full,yan2018tuning} with the exception of the cases when passivation species, in particular tetraphenylphosphonium halides (TPPX)~\cite{tong2023situ} and pentaerythritol tetrakis(3-mercaptopropionate) (PETMP)~\cite{huang2025surface}, are employed to heal surface defects and preserve PLQY.

\begin{figure*}[t!]
\centering
\includegraphics[width=0.65\columnwidth]{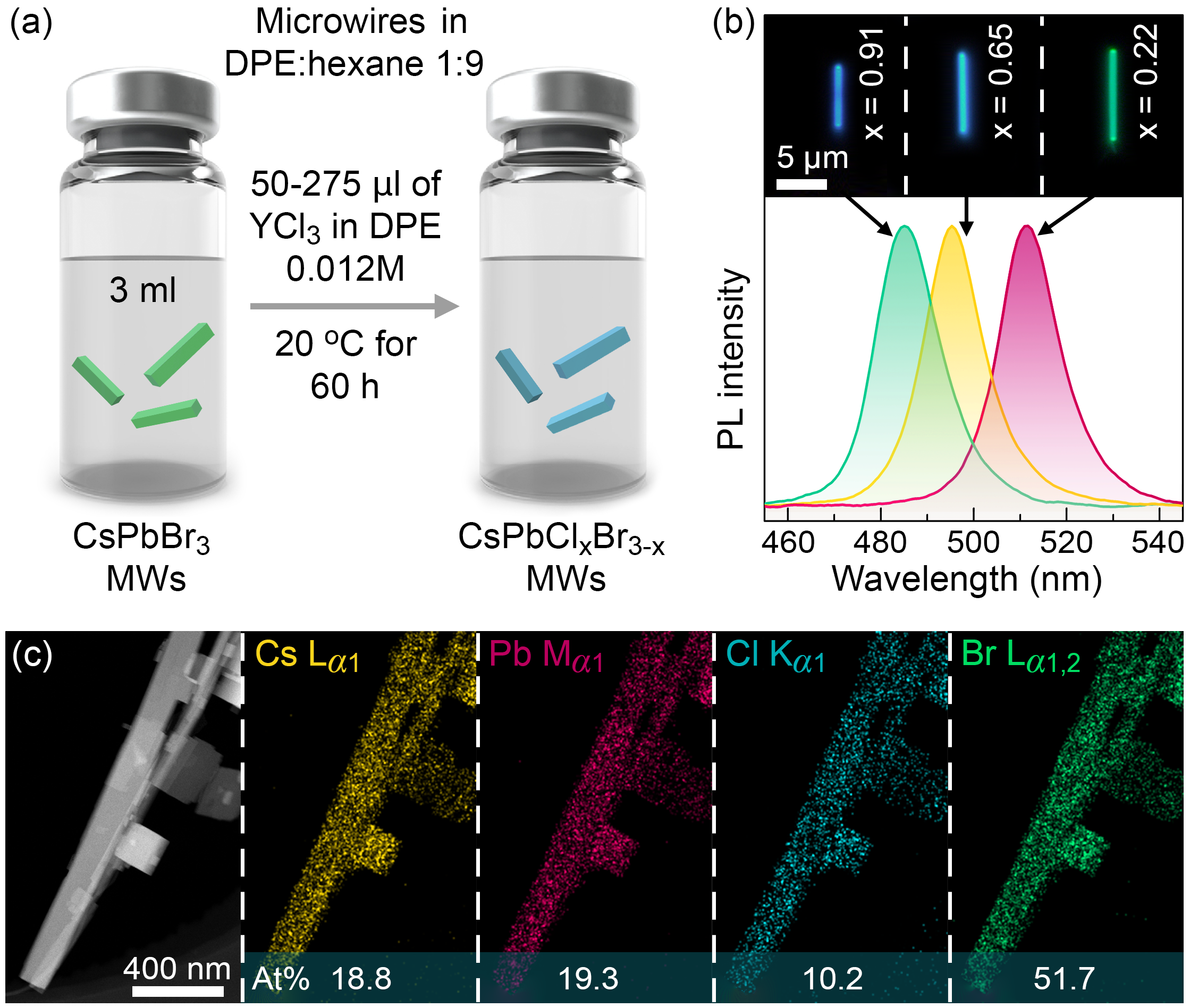}
\caption{(a) Schematic illustration of alloyed CsPbCl$_x$Br$_{3-x}$ MWs synthesis. (b) Fluorescence images and PL spectra of MWs with various content $x$ of chlorine in the perovskite lattice. (c) HAADF-STEM image and EDX mapping of CsPbCl$_{0.49}$Br$_{2.51}$ perovskite particles, revealing a uniform distribution of Cs, Pb, Cl, and Br elements.}
\label{fig4}
\end{figure*}

In our halide exchange reaction, we utilize YCl$_3$, which was proven to enhance PLQY of initially weakly luminescent CsPbCl$_3$ NCs by 60 times~\cite{ahmed2018giant}. This happens because of Y$^{3+}$ and Cl$^-$ ions efficiently occupy the surface ion vacancies, enriching the density of states in the conduction band (CB) without creating any midgap states~\cite{ahmed2018giant}. According to that, 50-275 $\upmu$L of YCl$_3$ 0.012 M solution in DPE is added to laser MWs in 3 mL of DPE:hexane 1:9 mixture (0.3 mL aliquot is mixed with 2.7 mL of hexane) and kept at 20 $^{\circ}$C for 60 h (Figure 4a). Obviously, a slow rate of the exchange reaction is explained by the halide ion diffusion in microscopic perovskite particles~\cite{lai2018intrinsic}. Furthermore, the reaction is concentration-limited by YCl$_3$, hence the exchange process, which we monitor by PL spectrum, slows down exponentially and the reaction mixture reaches its equilibrium (Figure S3). For 50 $\upmu$L additive, the resultant alloyed MWs have a mean wavelength of PL peak 515 nm, whilst 275 $\upmu$L additive yields MWs emitting at ca. 485 nm. Fluorescence images and PL spectra of CsPbCl$_x$Br$_{3-x}$ MWs (x is determined by PL peak energy~\cite{liashenko2019electronic}) show homogeneous luminescence and single peak emission, respectively, meaning uniform distribution of halide ions on scales $>$ 250 nm (Figure 4b). EDX mapping confirms the complete mixing of Cl$^-$ and Br$^-$ ions in perovskite particles on the nanoscale (Figure 4c). No noticeable signal of Y$^{3+}$ is detected that implies its species interact with MWs surface only to mediate the halide exchange.

\begin{figure*}[t!]
\centering
\includegraphics[width=0.75\columnwidth]{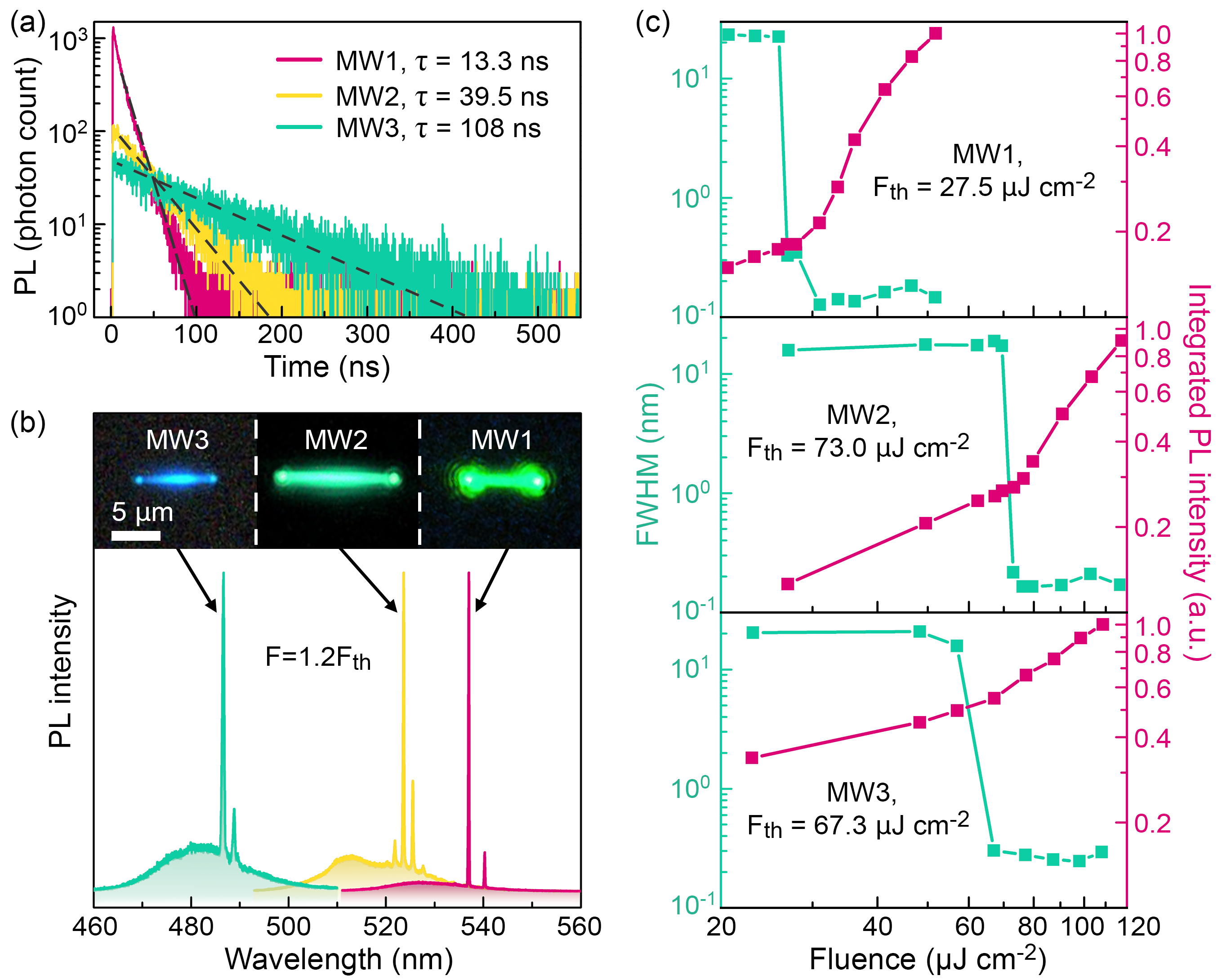}
\caption{(a) TRPL kinetics for MWs for pure bromine and mixed halide MWs, where each of them was measured for the same amount of time t = 110 s. The inset image demonstrates normalized PL decay kinetics. (b) Photoluminescence spectra of MWs at excitation fluence equal to 1.2 of laser generation threshold fluence F$_{th}$. (c) Dependencies of PL intensity and FWHM value on the pumping fluence.}
\label{fig5}
\end{figure*}

To estimate how the alloying impacts the optical properties of MWs, we carry out time-resolved photoluminescence (TRPL) measurements. For this purpose, MWs\textbf{1}-\textbf{3} exhibiting PL peak at 525, 512, and 482 nm, respectively, are excited by 405 nm laser pulses with pulse duration $\tau$ = 200 fs at repetition frequency f = 100 kHz. Excitation fluence is set up to 100 nJ cm$^{-2}$ to achieve monomolecular trap-assisted radiative recombination in the pure bromide NW\textbf{1}. Figure 5a shows PL decay curves acquired for the same time interval. One can see how alloying drops PL intensity down, however, increases its lifetime: $\tau$ = 13.3 ns for NW\textbf{1}, 39.5 ns for NW\textbf{2}, and 108 ns for NW\textbf{3}. Taking into account a minor difference in extinction (see band-to-band absorption in SI~\cite{liashenko2019electronic}) between MW\textbf{1} (CsPbBr$_3$, $\alpha\approx$ 0.5$\times$10$^5$ cm$^{-1}$) and MW\textbf{3} (CsPbCl$_{0.9}$Br$_{2.1}$, $\alpha\approx$ 0.4$\times$10$^5$ cm$^{-1}$), it is clear that the same fluence can not produce a major difference in the density of photogenerated carriers. Therefore, the drop of PL is related to the generation of midgap trap (defect) states. Concurrently, the generated shallow trap states contribute to the elongation of lifetime through carrier trapping and detrapping events~\cite{jin2020s}. To evaluate the relative change in PLQY for MWs\textbf{1}-\textbf{3}, we compare areas under the decay curves and conclude that PLQY decreases by 2.6 times for MW\textbf{2} and 2.8 times for MW\textbf{3}. Still, this deterioration is not as dramatic to suppress lasing.

Lasing in the NWs\textbf{1}-\textbf{3} is induced by 350 nm pulsed laser excitation ($\tau$ = 150 fs, f = 10 kHz). Optical images of the MWs pumped with fluence above the laser threshold value (F$_{th}$) and their spectra are shown in Figure 5b. The spectra consist of 2-3 intensive sharp lines assigned to FP modes along with broadband spontaneous emission (SE). The contribution of SE to the entire photoluminescence increases with chlorine content $x$ and is consistent with the TRPL data, since the residual SE signal comes from the delayed radiative recombination of detrapped carriers.

The reduced PLQY and residual SE influence the laser threshold value, FP modes linewidth ($\delta \lambda$), and the shape of S-curve (`pump-power' dependence) for alloyed MWs \textbf{2} and \textbf{3} (Figure 5c). Indeed, F$_{th}$ value tends to increase from 27.5 to ca. 70 $\upmu$J cm$^{-2}$ as PLQY goes down, whereas maximum Q-factor (Q =$\lambda_{mode}$/$\delta \lambda$) of the dominant FP mode descends as follows: 3030 for MW\textbf{1}, 2490 for MW\textbf{2}, and 1080 for MW\textbf{3}. At the same time, the S-curve is getting less steep because of the greater contribution of SE to integrated PL intensity (Figure 5c).

\subsection{Optoelectronic integrated device}
The obtained colloidal MWs are utilized for the engineering of a device consisting of CsPb(Cl, Br)$_3$ microlaser sitting on the $\alpha$-MoO$_3$ nanowaveguide and CsPbBr$_3$ MW photodetector on gold electrodes. Nanowaveguides are synthesized by sublimation of $\alpha$-MoO$_3$ powder on Si/SiO$_2$ substrate (Figure S4) according to the protocol described in the Experimental Section. Their superior crystal quality is proven by HRTEM images revealing a directional growth of orthorhomic lattice (space group Pbnm~\cite{wooster1931crystal}) along [001] axis, low surface roughness (less than 2 nm), regular crystal faces, and sharp spots observed in selected area electron diffraction (Figure S5). The absence of the submicron crystallites, which can scatter the light traveling along the nanowaveguide, is verified by a dark-field image measured in the transmission mode (Figure S6). We apply a polymer-assisted dry transfer procedure to 84 $\upmu$m single nanowaveguide~\cite{matchenya2025short} to align it along the planar electrodes (Figure S7) fabricated on a glass substrate by photolithography (for details see Experimental Section). Thereafter, the perovskite MWs drop-casted on a glass slide and rinsed with hexane to wash away the traces of DPE are subjected to the same dry transfer procedure and subsequently integrated with the electrodes (17.8 $\upmu$m tribromide MW) and the nanowaveguide (5 $\upmu$m mixed-halide MW). The entire device is illustrated in Figure 6a. Atomic force microscopy (AFM) images (Figure 6b) reveal the following dimensions of the produced design: i) MW photodetector possesses 1.38$\times$2.1 $\upmu$m cross-section (1.47 $\upmu$m height in Figure 6b includes an air gap due to 90 nm gold electrodes); ii) nanowaveguide exhibits a rectangular cross-section 0.175$\times$1.2 $\upmu$m; iii) MW laser dimensions equal to 5$\times$0.84$\times$1.3 $\upmu$m. Pumping the microlaser by 400 nm laser pulses ($\tau$ = 250 fs, f = 500 Hz) at fluence below the threshold F$_{th} = $~0.75 mJ cm$^{-2}$ gives spontaneous emission (SE), whilst surpassing this value results in multimode lasing (Figures 6c,d). For the explanation of the high threshold value, see Experimental Section. Note laser modes of CsPb(Cl,Br)$_3$ microlaser appear at ca. 500 nm and overlap well with the absorption spectrum of CsPbBr$_3$, which enables efficient generation of photoexcited carriers in the latter. At the same time, lasing occurs in the spectral range of the maximum transmission of the nanowaveguide (Figure 6d).

To examine the functionality of the produced device, we synchronize electrical and optical stimuli: fs laser pulses come to the MW laser in the middle of 1 ms rectangular voltage pulses applied to the MW photodetector. The reason for the selection of such a synchronization mode is the demonstration of a possible photocurrent memory effect invoked by long-living electron traps~\cite{marunchenko2024charge, marunchenko2024hidden} excited by intense ps pulses from the microlaser~\cite{zhu2015lead}. Energy diagrams describing the device operating regimes are shown in Figure 6e. 

\begin{figure*}[t!]
\centering
\includegraphics[width=1\columnwidth]{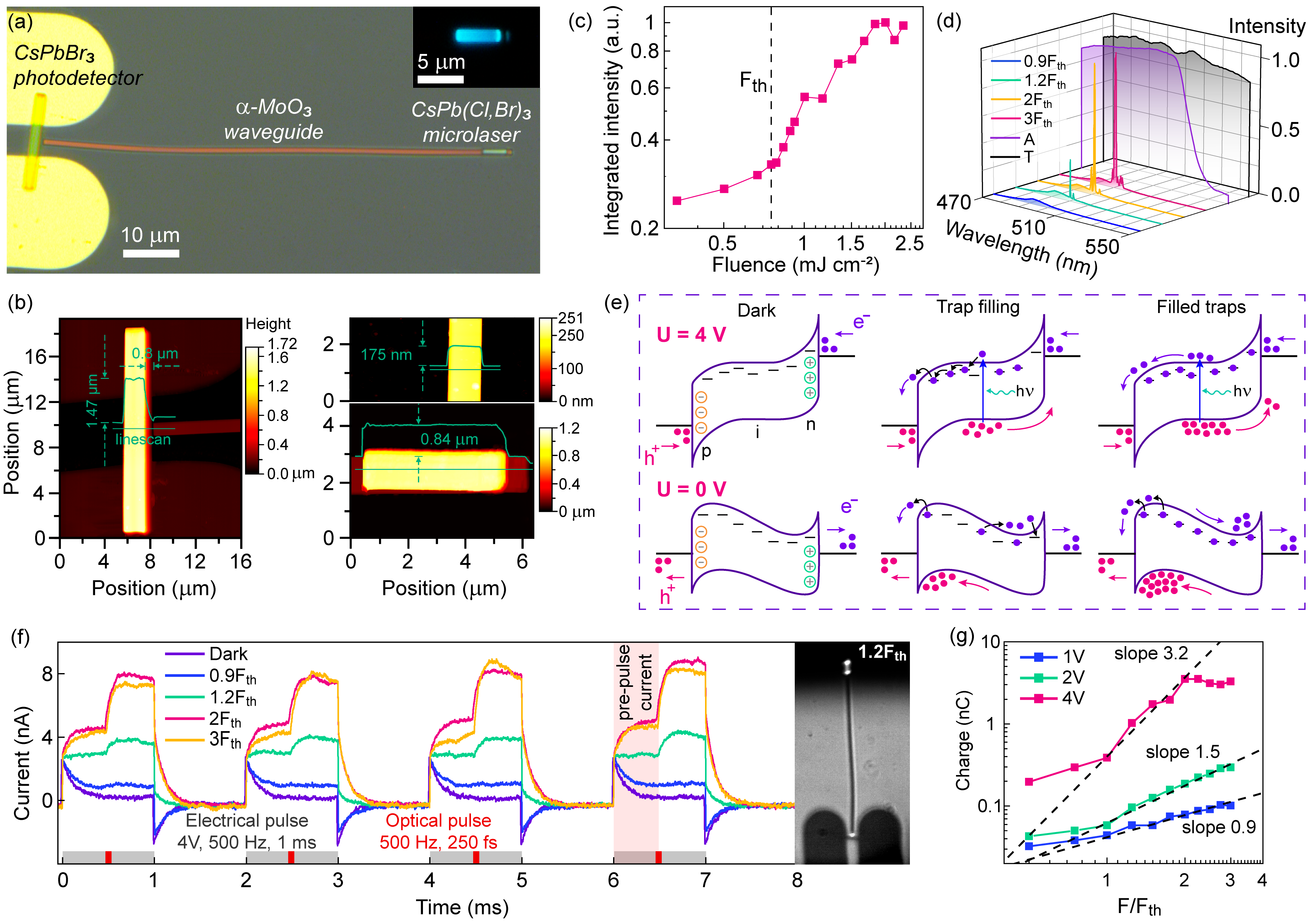}
\caption{(a) Optical image of the integrated optoelectronic device consisting of CsPbBr$_3$ MW photodetector on gold electrodes, $\alpha$-MoO$_3$ nanowaveguide, and CsPb(Cl,Br)$_3$ MW laser sitting on the nanowaveguide. The inset image shows PL of the mixed-halide microcrystal. (b) AFM images of different parts of the device. Green solid lines represent profiles along the linescans. Left image visualizes a 0.8 $\upmu$m gap between the MW photodetector and the nanowaveguide. The top right image exhibits the middle part of the nanowaveguide. The bottom right image shows MW laser atop the nanowaveguide and oriented along its axis. (c) `Pump-power' dependence for the MW laser demonstrating a lasing threshold F$_{th}$ = 0.75 mJ cm$^{-2}$. (d) Optical properties of the device's constituents: PL spectra of the MW laser below and above the threshold, normalized absorption spectrum of the MW photodetector, and normalized transmission spectrum of the nanowaveguide. (e) Energy band diagrams describing three operation regimes (dark mode, trap filling, and filled traps) of the biased and unbiased integrated device. (f)  Optoelectronic responses of the device subjected to hybrid pulsed stimuli at various optical pump fluences. The side image illustrates the device operating under F = 1.2F$_{th}$. One can see how laser light decouples from the end facet of the nanowaveguide. (g) The dependence of the electric charge  generated by the MW photodetector on F/F$_{th}$ value. The nonlinearity order (slope) above the lasing threshold increases with bias.}     
\label{fig6}
\end{figure*}

Under dark conditions, the device demonstrates charging and discharging currents typical for a capacitor (Figure 6f). When the microlaser operates below the threshold (F = 0.9F$_{th}$), the photodetector capacity-like response is complemented by a weak photocurrent. Interestingly, this weak current persists and can be observed before the optical pulse strikes the photodetector. We call it a `pre-pulse' current. As the microlaser surpasses the lasing threshold, much more light decouples from the end facet of the nanowaveguide (see the side image in Figure 6f). In this regime, the `pre-pulse' current flowing through the MW photodetector increases with the intensity of the input light and is high enough to mask the charging current. Along with that, we detect a step-like jump of the photocurrent which occurs simultaneously with the incoming optical pulse from the microlaser. Importantly, this photocurrent decays slowly after the end of the electrical pulse, which drastically modifies the falling edge of the optoelectronic response pulse as compared to a capacitor-like response.

According to the energy band diagrams (Figure 6e), in the external electric field, mobile halide vacancies~\cite{senocrate2019solid,tyagi2025tracing} move towards a negative electrode and afford a lightly doped n-type region. Concurrently, the negatively charged halide interstitials give a lightly doped p-type region at the positive electrode side. The lightly doped regions yield the band bending at the perovskite-gold interfaces. The produced energy barriers are too high and thick for efficient carrier injection. Therefore, under electrical excitation only, the device behaves like a capacitor: it demonstrates decay of positive and negative currents at the beginning and the end of a voltage pulse, respectively (dark current mode in Figure 6e). 

Below the lasing threshold, a weak optical pulse comes to the MW photodetector and generates carriers. Photoexcited electrons occupy shallow traps located close to the conduction band minimum (CBM), whilst holes move freely in the valence band. Consequently, a hole diffusion current contributes mainly to the photocurrent of the biased device (trap filling mode in Figure 6e). In the unbiased device, a weak current, which slightly compensates the discharging current, occurs owing to carriers detrapped at the perovskite-electrode interface. Along with that, there are carriers accumulated in the wells behind the Schottky barriers. These carriers, in particular holes, afford the `pre-pulse' current when the next electrical pulse comes to the MW photodetector and unblocks the holes. 

Above the lasing threshold, the intensity of PL outcoupled from the end facet of the nanowaveguide is sufficient for reaching the high concentration of electrons to fill all the traps (filled traps regime in Figure 6e). This results in an excess of free electrons in the conduction band, which contribute to a step-like rise in photocurrent level. The unbiased device, in turn, shows a pronounced decay of the photocurrent due to the release of electrons from numerous traps at the perovskite-electrode interface.

To quantitatively demonstrate the dependence of the optoelectronic response on external stimuli, we plot the charge (integrated current) versus F/F$_{th}$ on a log-log scale at various biases (Figure 6g). One can see that above F$_{th}$ all the curves change their slope and exhibit growth, which can be approximated by 0.9, 1.5, and 3.2 orders of nonlinearity for biases of 1 V, 2 V, and 4 V, respectively. We underline a complex nature of the observed nonlinearity which relies on following phenomena: i) a transition from spontaneous PL to lasing in the MW laser causes nonlinearity in the generation of carriers contributing to the diffusion current in the MW photodetector; ii) in the unbiased device, an anode collects electrons released from shallow traps (volatile memory); iii) switching from the unbiased to the biased device grants access to carriers blocked by Schottky barriers (non-volatile memory). Taking all of these into account, we propose our device as a new platform for on-chip neuromorphic computing, which is an advanced follow-up of the recently established one based on a bare perovskite exciton-polariton microlaser~\cite{opala2025perovskite}.

\section{Conclusion}
\label{conclusion}
In summary, the proposed approach to the colloidal synthesis of CsPbBr$_3$ microwires (MWs) demonstrates a significant advancement in the field of functional materials for state-of-the-art optoelectronic devices. The developed methodology, based on the use of a diphenyl ether coordinating solvent, reduced excess of organic ligands, and controlled nanocrystal regrowth, enables the production of high-quality 5--20 $\upmu$m MWs with excellent crystallinity and decent optical and electrical properties required for lasing and light photodetection, respectively. We have provided a detailed insight into the colloidal particles evolution from nanocrystals to nanowires and subsequently to microwires, offering a deeper understanding of the mechanisms that govern self-assembly and growth in perovskite systems. Furthermore, the ability to finely tune the optical properties of MWs through a post-synthetic YCl$_3$-assisted ion exchange has been established. As a result, alloyed CsPb(Cl,Br)$_3$ MWs with tailored lasing in the 485--540 nm range have been obtained.

We have leveraged the simple processing of MWs and their tunable optical properties to engineer a miniature optoelectronic device comprising a microlaser, a nanowaveguide, and a microphotodetector. This device has exhibited a nonlinear response to hybrid optical-electrical stimulation that has potential applicability in neuromorphic systems. Thus our findings not only expand the fundamental understanding of the synthesis and properties of perovskite microstructures but also establish a solid foundation for future advances in integrated optoelectronics and emerging computational technologies.	

\clearpage

\section{Experimental Section}
\subsection*{Materials}
 Cesium carbonate (Cs$_2$CO$_3$, Sigma-Aldrich, 99.995$\%$ trace metals basis), lead(II) bromide (PbBr$_2$, Alfa Aesar, 99.998~$\%$), 1-octadecene (ODE, Sigma-Aldrich, technical grade, 90$\%$), oleic acid (OA, Sigma-Aldrich, technical grade, 90$\%$), oleylamine (OlAm, Acros Organics, technical grade, 90$\%$), diphenyl ether (DPE, Sigma-Aldrich, $>$99$\%$), n-hexane (Acros Organics, for analysis, 95$\%$), acetone (Vekton), molybdenum(VI) oxide (MoO$_3$, Sigma-Aldrich, 99.97$\%$ trace metals basis), silicone elastomer and curing agent (Dow Corning Sylgard 184), silicon/silicon dioxide wafers (Si/SiO$_2$, Ossila, 300 nm oxide), gold target (Au, Girmet, 99.99$\%$ ), AZ 1505 (MicroChemicals), LOR 5B (Kayaku Advanced Materials) were used as received.

\subsection*{Synthesis of cesium oleate solution}

Cs$_2$CO$_3$ (0.814 g), 1-octadecene (ODE, 40 ml), and oleic acid (OA, 2.5 ml) were loaded into a 100 ml flask, 
dried under vacuum at 120 $^o$C for 1 h, and then heated up to 150 $^o$C under stirring and N$_2$ protection, yielding a clear cesium oleate (CsOA) 0.125 M precursor solution. The solution was cooled down to room temperature for storage, and preheated up to 120 $^o$C before use.

\subsection*{Synthesis of MoO$_3$ nanowaveguides}

MoO$_3$ nanowaveguides were grown from MoO$_3$ powder using a high-temperature titanium hotplate (Harry Gestigkeit). Two Si/SiO$_2$ substrates (10$\times$10 mm) were cleaned in an ultrasonic bath in acetone for 5 min, followed by isopropanol and deionized water, and then dried with a nitrogen gun. Subsequently, 7 mg of MoO$_3$ powder was placed on the bottom Si/SiO$_2$ substrate and spread over an area of 0.25 cm$^2$. Silicon spacers (1 mm thick) were used to separate the bottom substrate from a top substrate placed upside down. The entire assembly was positioned on a hotplate preheated to 580 $^\circ$C and maintained at this temperature for 8 h. After growth, the top substrate was removed from the hotplate without allowing it to cool. As a result, MoO$_3$ hedgehog-like structures consisting of nanowaveguides with lengths of up to 100 $\upmu$m were obtained on this substrate (Figure S4).

\subsection*{Structural and basic optical measurements}
XRD patterns for particles obtained from the aliquots were measured using a SmartLab diffractometer (Rigaku) equipped with
a 9 kW rotating Cu anode X-ray tube. Conventional TEM and HAADF-STEM images, electron diffraction patterns, and  EDX maps were obtained on a Titan Themis Z transmission electron microscope (Thermo Fisher Scientific) operated at 200 kV. The absorption spectrum of the colloidal solution (sample 1) was recorded on a UV-3600 spectrophotometer (Shimadzu). The photoluminescence spectra of colloidal particles and MWs in DPE:hexane solution were recorded on a Cary Eclipse fluorescence spectrometer (Agilent). The bright-field and fluorescent images of the samples were obtained by using an Axio Imager A2m (Carl Zeiss) microscope with a 100$\times$ objective (Carl Zeiss EC Epiplan-NEOFLUAR). PL spectra of isolated MWs were recorded by using an optical fiber spectrometer (Ocean Optics QE Pro) coupled with the microscope in the fluorescent regime. The area of detection was a spot of 2~$\upmu$m diameter. The absence of the volume defects in MWs was verified by means of dark-field microscopy as follows: MWs were illuminated at an oblique angle 65$^o$ to the normal of the surface by s-polarized light from a halogen lamp (HL2000-FHSA) through a weakly focusing objective (Mitutoyo M Plan Apo NIR, 10$\times$, NA = 0.28). Scattered light was collected from the top by a 50$\times$ objective (Mitutoyo M Plan APO NIR, NA = 0.42) and sent to a CCD camera (Canon 400 D). The transmission spectra of the MWs were measured in the same geometry by using a LabRam HR spectrometer (Horiba).

\subsection*{Lasing and time-resolved PL measurements}
Laser generation in isolated MWs was induced by femtosecond laser excitation (TeMA, Avesta Project) with a wavelength $\lambda$ =~350 nm, pulse duration $\tau$ =~150 fs at pulse repetition rate f~=~10 kHz. The beam was focused on a sample by using a 20$\times$ objective (Mitutoyo M Plan APO NUV, NA = 0.42) at 30$^o$ angle. Emission signal was collected with 50$\times$ objective (Mitutoyo M Plan APO VIS, NA = 0.55) and processed by LabRam HR spectrometer with 1800 grooves/mm grating. Excitation light was blocked by a longpass filter (FELH 450, Thorlabs). 

To study PL decay in isolated MWs, they were excited by 405 nm laser pulses ($\tau$ = 200 fs, f = 100 kHz) at fluence F = 100 nJ cm$^{-2}$. The photoluminescence signal was transmitted through a fiber and sent to a single photon avalanche diode (PDM, Micro Photon Devices) connected to a time-correlated single photon counting system (PicoHarp 300, PicoQuant).

\subsection*{Fabrication of gold electrodes}

Electrical contacts were fabricated on 2$\times$2 cm glass substrate with a channel length (distance between contacts) of $\approx$ 7 $\upmu$m and a width of $\approx$ 620 $\upmu$m. The contacts pattern was defined using UV laser photolithography (Heidelberg Micro Pattern Generator $\upmu$PG 101) on double-layer photoresist (AZ1505 positive photoresist onto LOR5B sublayer). A 90 nm gold layer was then deposited by the Quorum Q150T ES magnetron sputtering, followed by a lift-off process in acetone (Figure S7).

\subsection*{Measurements of the integrated device morphology and optoelectronic response}

AFM images of the device constituents were obtained using a Bruker Multimode V8 in Quantitative NanoMechanical Mapping PeakForce tapping mode. The force constant of the cantilever was set to 3.6 N m$^{-1}$, the resonant frequency was around 77 kHz, the thickness and width of the cantilever were 3 $\upmu$m and 34 $\upmu$m, respectively. For scanning, we employed a scan rate of 0.4 Hz and a setpoint force of 10 nN.

For the optoelectrical characterization of the integrated device, we developed an experimental setup capable of simultaneous real-space imaging and spectral tracking of laser emission (Figure S8). Spectral data were acquired using a Shamrock SR-500 spectrometer coupled with an Andor iXon Ultra 888 scientific camera with spectral resolution of 0.05 nm. Optical excitation of the CsPb(Br, Cl)$_3$ microlaser was provided by a Coherent Libra femtosecond laser operating at 800 nm, with a pulse width of 250 fs and a repetition rate of 500 Hz, passed through a second-harmonic generator (SHG). Note that the lasing threshold was about 1 order of magnitude higher than obtained on the TeMA laser (Avesta Project) at 10 kHz repetition rate. Such a difference is related to the concentration of hidden photoexcitations - electrons residing in shallow traps~\cite{marunchenko2024hidden}. Pumping the MW at 10 kHz establishes a quasi-equilibrium state with partially filled traps and consequently requires lower fluence to induce lasing compared to pumping at 500 Hz, which leads to complete trap depopulation.            

The resulting emission from the perovskite microlaser was guided through a MoO$_3$ waveguide and subsequently detected by a CsPbBr$_3$ photodetector. To probe the photodetector response, a series of 1 ms rectangular electrical pulses was applied using a Keysight 33500B Series waveform generator. These pulses were synchronized with the optical excitation at a matching repetition rate of 500 Hz. The phase between the optical excitation and the electrical pulses was adjusted to ensure the laser pulse arrived at the temporal center of the electrical square pulse. Finally, the electrical signal generated by the perovskite photodetector was recorded using a Keysight InfiniiVision DSOX3054T oscilloscope.

\section*{Supporting Information} \par
Supporting Information is available from the Wiley Online Library or from the author.

\section*{Acknowledgements}
This work was supported by the Russian Science Foundation (project No. 24-73-10072, \url{https://rscf.ru/prjcard\_int?24-73-10072}). TEM, HAADF-STEM images and EDX maps were measured in the Advanced Imaging Core
Facility (AICF) of Skolkovo Institute of Science and Technology (Skoltech). The authors thank Dr. Adrian Zavodov for the TEM measurements of MoO$_3$ nanowaveguides. 

\section*{Author contribution}

A.P. originated the idea. E.S., D.S., D.T., G.V., and A.P. synthesized perovskite microwires (MWs). E.S. and D.S. studied the optical properties of colloidal particles and MWs. D.T., G.V., N.K., M.K., and A.P. studied the morphology and structural properties of colloidal particles and MWs. Y.M. synthesized MoO$_3$ nanowaveguides and studied their structural properties. A.K., I.M. S.B., and A.P. fabricated the optoelectronic device and measured its optoelectronic response. V.L. measured and analyzed the surface morphology of the device. I.M., Y.T., and A.P. contributed to the writing of the original draft. A.P. supervised the project. All authors contributed to the discussion and the writing of the manuscript.

\section*{Conflict of Interest}
The authors declare no conflict of interest.

\section*{Data Availability Statement}
The data that support the ﬁndings of this study are available in thesupplementary material of this article.

\clearpage

\medskip

%

\bibliographystyle{MSP}
\bibliography{biblio}


\newpage

\begin{figure}
  \centering
\textbf{Table of Contents}\\

\medskip
  \includegraphics{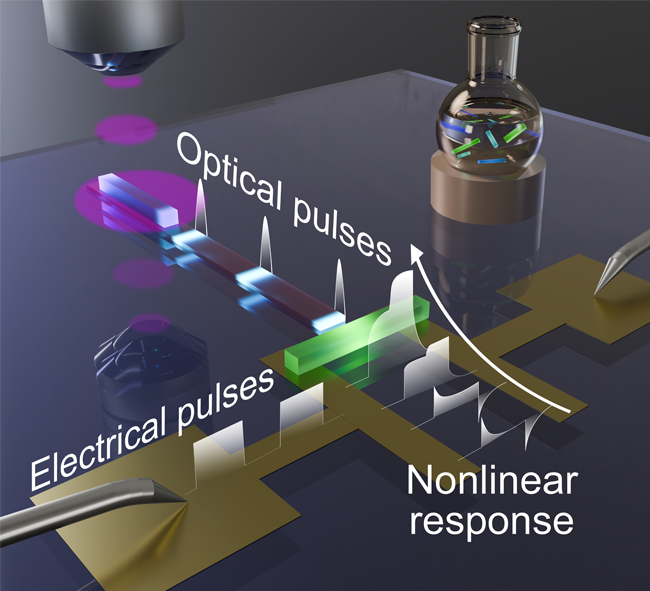}
  \medskip
  \caption*{We present the colloidal synthesis of perovskite microwires exhibiting spectrally tunable lasing and photodetection. This platform enables the integration of microscale devices that demonstrate nonlinear responses to combined optical and electrical stimulation, highlighting their potential for neuromorphic computing.}
\end{figure}

\end{document}


\pagestyle{fancy}

\newcommand{\red}{\textcolor{red}}
\newcommand{\blue}{\textcolor{black}}

\renewcommand{\thefigure}{S\arabic{figure}}	
	\renewcommand{\theequation}{S\arabic{equation}}	
	\renewcommand{\thesection}{S\arabic{section}.}

\title{\center{\textbf{Supplementary Information}}
\\
Colloidal Nanocrystals Regrowth-Assisted Synthesis of Perovskite Microwire Lasers for Integrated Optoelectronics} 

\maketitle


\author{Elizaveta V. Sapozhnikova\textsuperscript{1,2$\dagger$},}
\author{Ivan A. Matchenya\textsuperscript{1$\dagger$},}
\author{Dmitry A. Tatarinov\textsuperscript{1,2},}
\author{Grigorii A. Verkhogliadov\textsuperscript{1},}
\author{Dmitry A. Semyonov\textsuperscript{2},}
\author{Maria A. Kirsanova\textsuperscript{1},}
\author{Natalia K. Kuzmenko\textsuperscript{2},}
\author{Julia S. Mironova\textsuperscript{1},}
\author{Arina O. Kalganova\textsuperscript{1},}
\author{Valeriya M. Levkovskaya\textsuperscript{1},}
\author{Stepan A. Baryshev\textsuperscript{1},}
\author{Yuxi Tian\textsuperscript{3},}
\author{Anatoly P. Pushkarev\textsuperscript{1}*}

\begin{affiliations}
$^1$ Skolkovo Institute of Science and Technology, Bolshoy Boulevard 30, bldg. 1, Moscow, 121205, Russia.

$^2$ ITMO University, Kronverksky Pr. 49, bldg. A, St. Petersburg, 197101, Russia.

$^3$ Key Laboratory of Mesoscopic Chemistry of MOE, State Key Laboratory of Analytical Chemistry for Life Science, School of Chemistry and Chemical Engineering, Nanjing University, 210023 Nanjing, China.

$^{\dagger}$These authors contributed equally to this work.

Anatoly P. Pushkarev

Email Address: an.pushkarev@skoltech.ru

\end{affiliations}

\makeatletter
\renewcommand{\@maketitle}{%
{%
\thispagestyle{empty}%
\vskip-36pt%
{\raggedright\sffamily\bfseries\fontsize{20}{25}\selectfont \@title\par}%
\vskip10pt
{\raggedright\sffamily\fontsize{12}{16}\selectfont  \@author\par}
\vskip25pt%
}%
}%
\makeatother

\begin{figure}[h]
    \centering
    \includegraphics[width=0.5\linewidth]{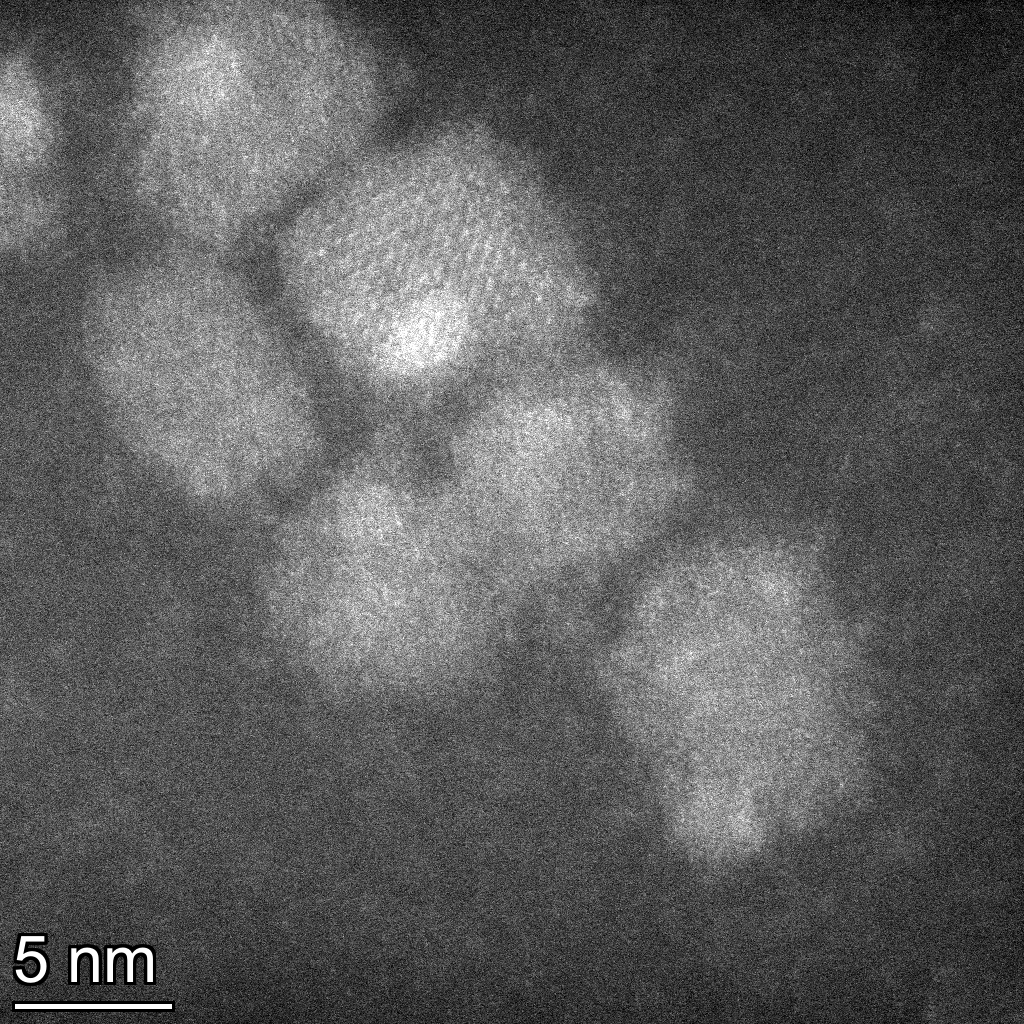}
    \caption{Small CsPbBr$_3$ NCs as-formed right after the injection of the CsOA solution into the PbBr$_2$/ligands solution at 140 $^\circ$C.}
    \label{fig:electrodes}
\end{figure}

\begin{figure}[h]
    \centering
    \includegraphics[width=0.9\linewidth]{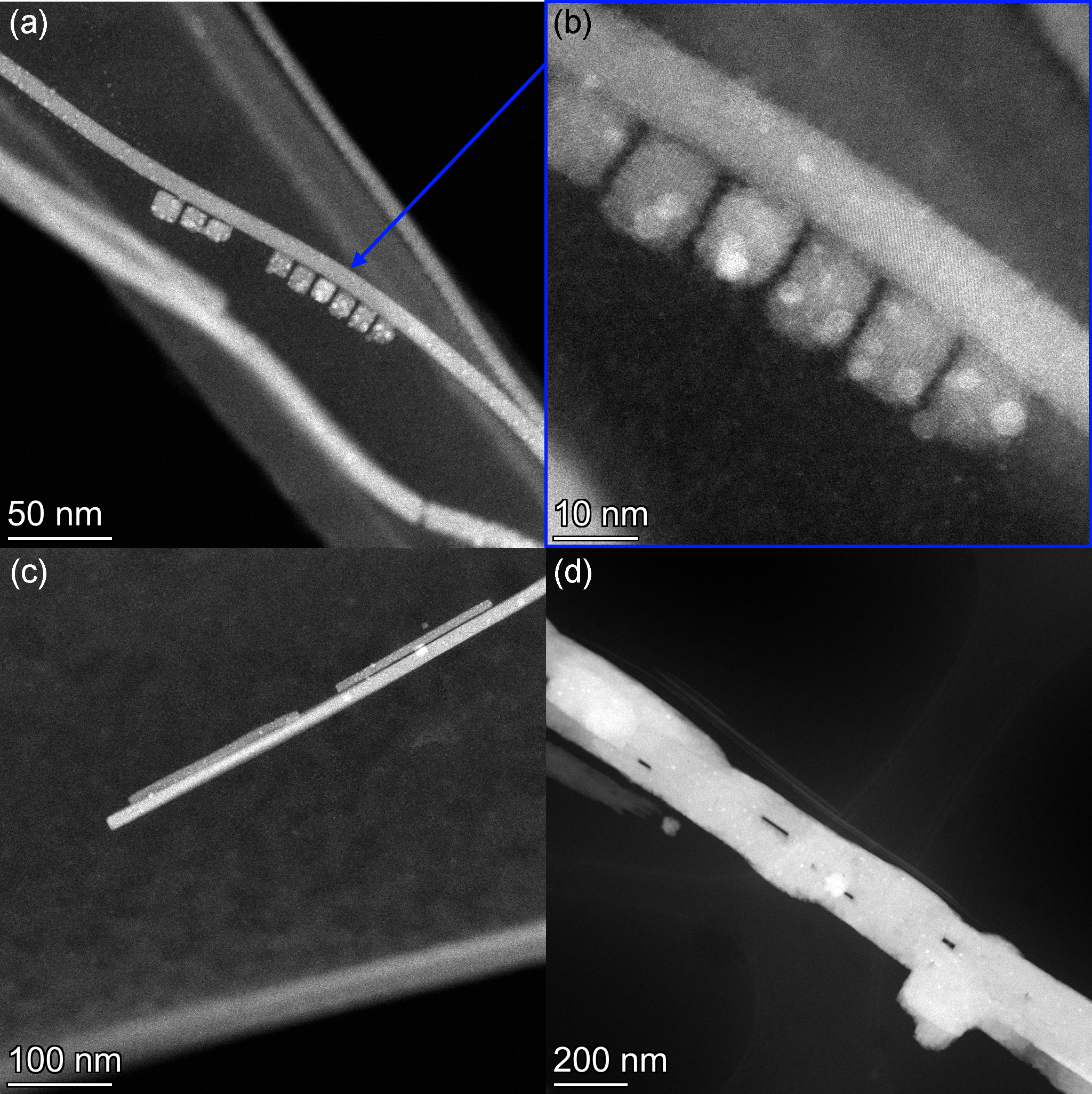}
    \caption{(a) Large-scale HAADF-STEM image showing the attachment of cubic-shaped CsPbBr$_3$ NCs to the (001) surface of a CsPbBr$_3$ NW. (b) High-resolution selected-area image revealing the merging of the NCs with the NW. (c) Coalescence of short NWs with a longer NW. (d) One-dimensional structure exhibiting rough surface morphology and voids formation resulting from the coalescence of particles of different sizes.}
    \label{fig:electrodes}
\end{figure}

\begin{figure}[h]
    \centering
    \includegraphics[width=0.9\linewidth]{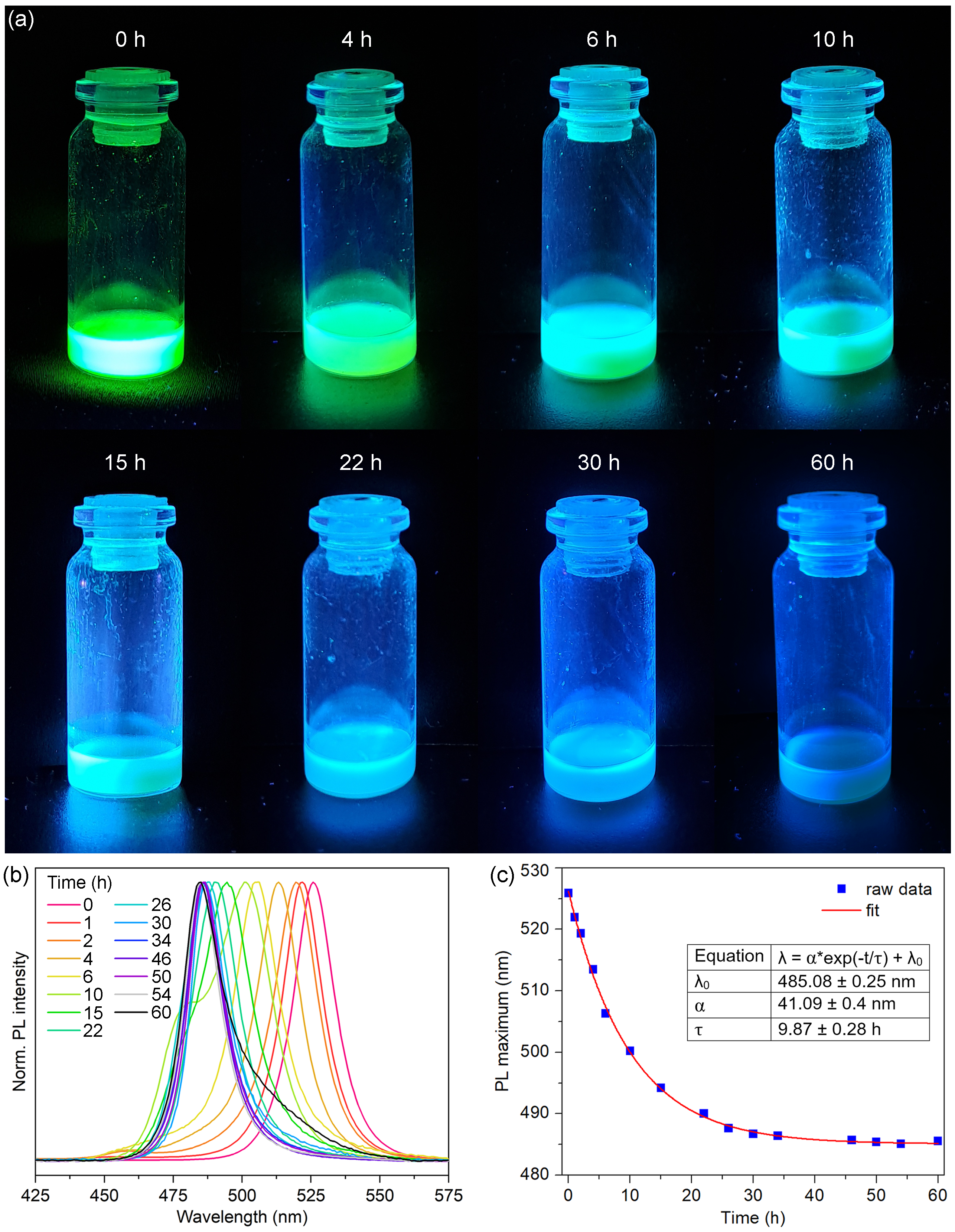}
    \caption{(a) Fluorescence visualization of slow halide exchange in CsPbBr$_3$ MWs stored in a DPE:hexane solution containing 275 $\upmu$L of YCl$_3$ in DPE (0.012 M). (b) PL spectral evolution. (c) Dependence of the PL peak wavelength on the duration of the halide-exchange reaction. The experimental data were fitted with an exponential decay function, with the fitting parameters listed in the inset Table.}
    \label{fig:electrodes}
\end{figure}

\begin{figure}[t!]
    \centering
   \includegraphics[width=0.6\linewidth]{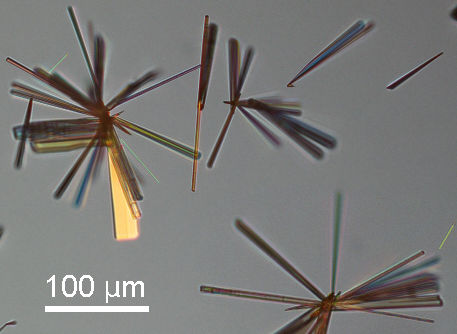}
    \caption{MoO$_3$ hedgehog-like structures on a Si/SiO$_2$ substrate, enabling the detachment of individual nanowaveguides using a PDMS lens.}
    \label{fig:electrodes}
\end{figure}

\begin{figure}[h]
    \centering
    \includegraphics[width=0.9\linewidth]{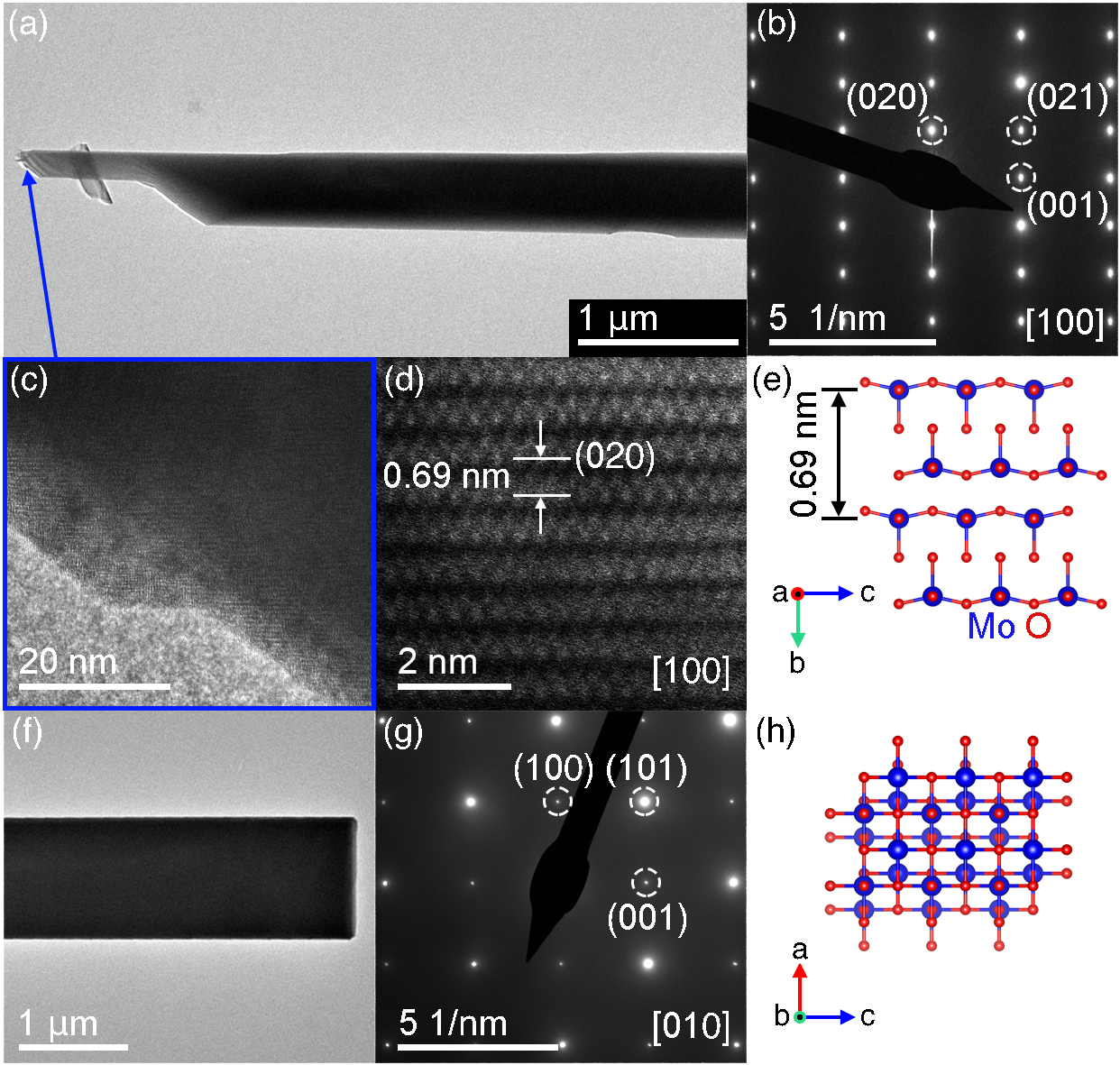}
    \caption{(a) Top-view TEM image of a MoO$_3$ waveguide featuring a sharp tip, enabling atomic-resolution imaging. The top facet of the waveguide corresponds to the (100) plane. (b) Selected-area electron diffraction (SAED) pattern acquired from the entire waveguide along the [100] zone axis, confirming crystal growth along the [001] direction. (c) High-resolution TEM image of the tip area.
(d) Visualization of the orthorhombic MoO$_3$ crystal lattice (space group Pbnm). (e) Correspondingly oriented reference crystal structure. (f) Regularly shaped end facet of a chipped waveguide, with the top facet corresponding to the (010) plane.
(g,h) SAED pattern of the waveguide along the [010] zone axis (g) and the reference crystal structure (h).}      
    \label{fig:electrodes}
\end{figure}

\begin{figure}[h]
    \centering
    \includegraphics[width=0.9\linewidth]{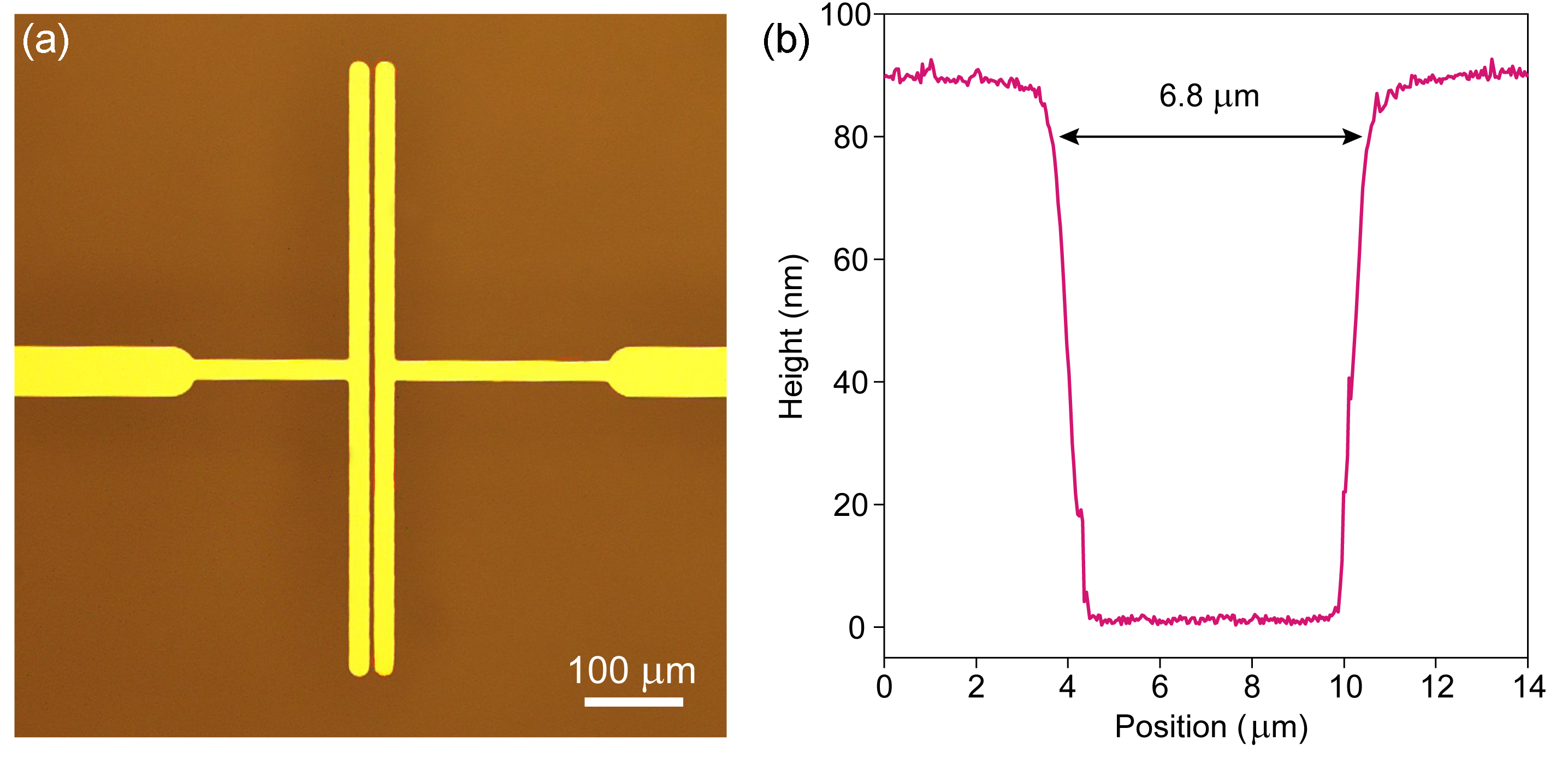}
    \caption{ (a) Bright-field image of gold electrodes on a glass substrate. (b) AFM profile across the electrode fingers, identifying their height of 90 nm and the interelectrode distance of approximately 6.8 $\upmu$.}
    \label{fig:electrodes}
\end{figure}

\begin{figure}[h]
    \centering
\includegraphics[width=0.4\linewidth]{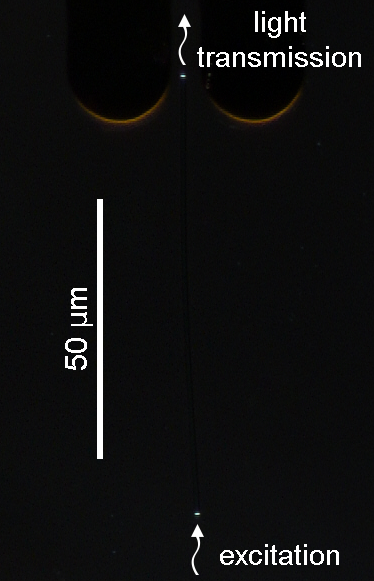}
    \caption{Dark-field transmission image of a MoO$_3$ nanowaveguide. Incident light illuminates one end facet, couples into the waveguide, propagates over a distance of 84 $\upmu$m without noticeable scattering, and outcouples from the opposite end facet.}
    \label{fig:electrodes}
\end{figure}

\begin{figure}[h]
    \centering
    \includegraphics[width=0.9\linewidth]{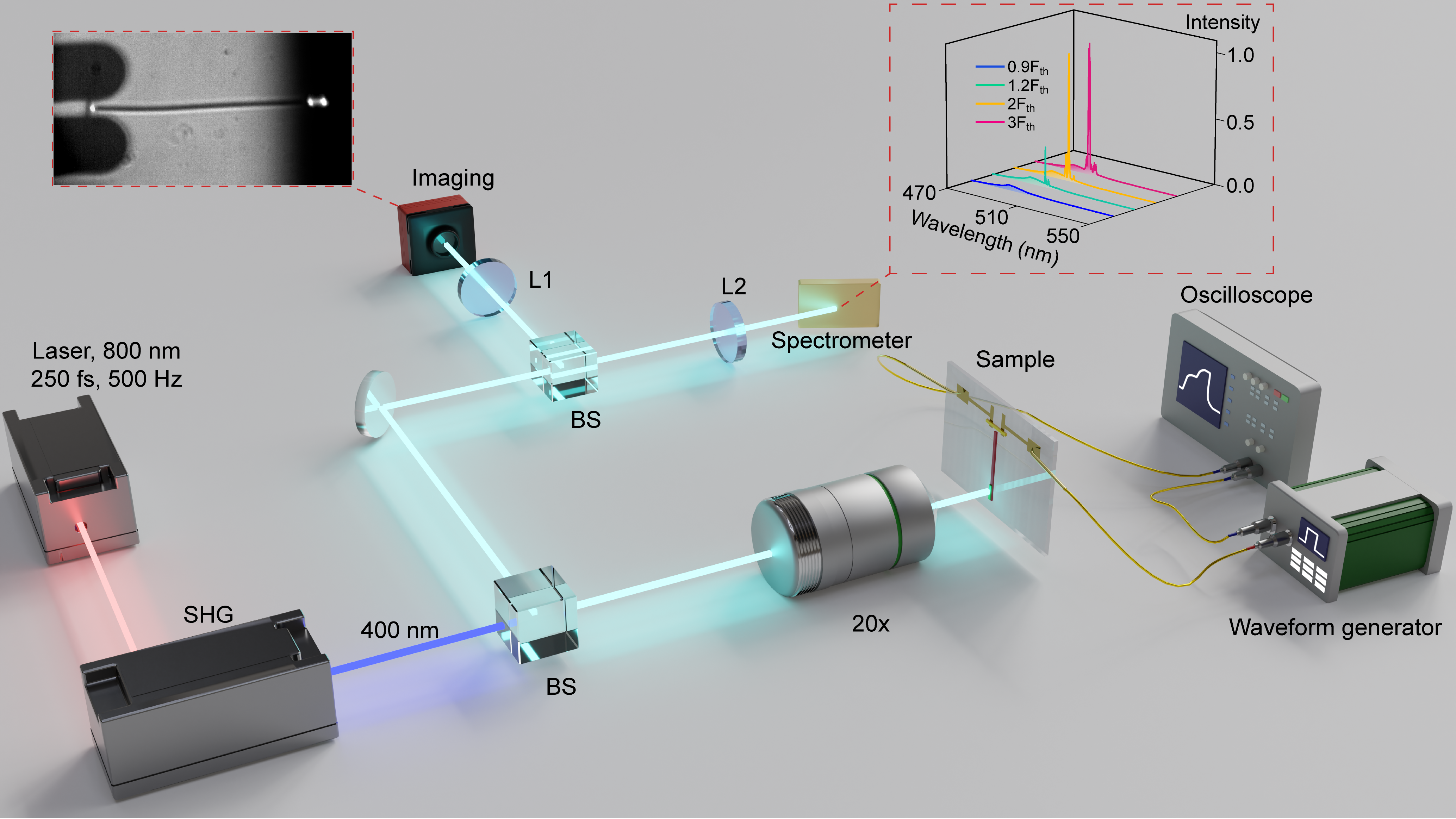}
    \caption{Experimental setup for the characterization of the optoelectronic response in the fabricated integrated device}
    \label{fig:setup}
\end{figure}